\documentclass[sigconf, natbib=true]{acmart}
\AtBeginDocument{%
  \providecommand\BibTeX{{%
    \normalfont B\kern-0.5em{\scshape i\kern-0.25em b}\kern-0.8em\TeX}}}

\setcopyright{acmcopyright}
\copyrightyear{2024}
\acmYear{2024}
\setcopyright{acmlicensed}\acmConference[SIGIR '24]{Proceedings of the 47th
International ACM SIGIR Conference on Research and Development in
Information Retrieval}{July 14--18, 2024}{Washington, DC, USA}
\acmBooktitle{Proceedings of the 47th International ACM SIGIR Conference on
Research and Development in Information Retrieval (SIGIR '24), July 14--18,
2024, Washington, DC, USA}
\acmDOI{10.1145/3626772.3657908}
\acmISBN{979-8-4007-0431-4/24/07}

\usepackage{tikz}
\usepackage{arydshln}
\usepackage[font=footnotesize,labelfont=bf,skip=3pt]{caption}
\usepackage{graphicx}
\usepackage{subcaption}

\begin{document}

\title[General-Purpose User Modeling with Behavioral Logs: A Snapchat Case Study]{General-Purpose User Modeling with Behavioral Logs}
\subtitle{A Snapchat Case Study}

\author{Qixiang Fang}
\authornote{Both authors contributed equally to this research.}
\email{q.fang@uu.nl}
\orcid{0000-0003-2689-6653}
\affiliation{%
  \institution{Utrecht University}
  \city{Utrecht}
  \country{Netherlands}
}

\author{Zhihan Zhou}
\authornotemark[1]
\email{zhihanzhou2020@u.northwestern.edu}
\affiliation{%
  \institution{Northwestern University}
  \city{Evanston}
  \state{Illinois}
  \country{USA}
}

\author{Francesco Barbieri}
\author{Yozen Liu}
\author{Leonardo Neves}
\affiliation{
  \institution{Snap Inc.}
  \city{Santa Monica}
  \state{California}
  \country{USA}}

\author{Dong Nguyen}
\author{Daniel Oberski}
\affiliation{%
  \institution{Utrecht University}
  \city{Utrecht}
  \country{Netherlands}}

\author{Maarten Bos}
\author{Ron Dotsch}
\affiliation{
  \institution{Snap Inc.}
  \city{Santa Monica}
  \state{California}
  \country{USA}}

\renewcommand{\shortauthors}{Qixiang Fang et al.}
\newcommand{\zhihan}[1]{{\color{red}{#1}}}
\newcommand{\qixiang}[1]{{\color{blue}{#1}}}

\begin{abstract}
Learning general-purpose user representations based on user behavioral logs is an increasingly popular user modeling approach. It benefits from easily available, privacy-friendly yet expressive data, and does not require extensive re-tuning of the upstream user model for different downstream tasks. While this approach has shown promise in search engines and e-commerce applications, its fit for instant messaging platforms, a cornerstone of modern digital communication, remains largely uncharted. We explore this research gap using Snapchat data as a case study. Specifically, we implement a Transformer-based user model with customized training objectives and show that the model can produce high-quality user representations across a broad range of evaluation tasks, among which we introduce three new downstream tasks that concern pivotal topics in user research: user safety, engagement and churn. 
We also tackle the challenge of efficient extrapolation of long sequences at inference time, by applying a novel positional encoding method. 
\end{abstract}




\maketitle

\section{Introduction}
\label{sec:intro}
Instant messaging (IM) apps, such as WhatsApp, WeChat and Snapchat, have become an indispensable part of modern lives~\cite{Bailey2016PerceptionsOM}. For businesses running these platforms, understanding their users' needs and delivering a superior experience is paramount. For example, with growing concerns about harassment, scams, and bots~\cite{Jain2021OnlineSN}, robust safety measures like timely identification and mitigation of malicious accounts are crucial. Furthermore, some IM apps (e.g., Snapchat and WeChat) suggest content (e.g., news, photos, videos) to users, requiring personalized recommendations.
To this end, many IM platforms leverage \textit{user modeling} techniques. 

User modeling concerns the process of creating representations of users that can capture and predict their actions, preferences, or intentions~\cite{Fischer2001UserMI}.
In this process, various data sources may be used, including user attributes~\cite{zhang_general-purpose_2020}, social media posts~\cite{andrews_learning_2019, chen_forum_2018}, and social networks~\cite{chen_forum_2018, modell_graph_2021, waller_generalists_2019}.  
An increasingly recognized alternative is \textit{user behavioral logs}, which are records of high-resolution, low-level events triggered by user actions in an information system~\cite{abb_reference_2022}. Examples are swiping left, viewing a video, and sending a text.

Using user behavioral logs for user modeling has many benefits. For instance, these logs contain rich and complex patterns, providing the basis for nuanced user representations. Also, because users tend to display consistent behaviors over time~\cite{chowdhury_ceam_2021}, behavioral logs remain more similar for the same user at different time points than between between two different users. This helps a model to learn personalized representations.  
Furthermore, the need for active user input like surveys is eliminated, thus reducing user burden. 
Lastly, behavioral logs are more privacy-friendly, unlike demographics, user-generated content (e.g., texts, images, videos), and locations, which users often consider private and sensitive~\cite{Pew2018, Fra2020, martin_what_2020, gilbert_measuring_2021}. 

For brevity, we refer to user modeling approaches that leverage user behavioral logs as \textbf{BLUM} (\textbf{B}ehavioral \textbf{L}og-based \textbf{U}ser \textbf{M}odeling).
BLUM can be either \textit{task-specific} or \textit{General-purpose}. In task-specific BLUM, separate user models are trained for different downstream tasks (e.g.,~\cite{li_multi-interest_2019, yu_adaptive_2019, zhou_deep_2018, Yang2018IKY, Liu2019CharacterizingAF, Baten2023PredictingFL, chen2022enhancing, cao2022sampling}). 
In contrast, in \textit{General-purpose} BLUM (\textbf{G-BLUM}), a primary, upstream user model learns general-purpose user representations not specific to a downstream task through some general training objectives~\cite{zhang_general-purpose_2020}. Then, user representations are extracted from this upstream model and used as direct input in downstream tasks, which often use lightweight learners (e.g., linear regression~\cite{chen_predictive_2018, chu_simcurl_2022}, logistic regression~\cite{chen_predictive_2018} and multilayer perceptron~\cite{zhang_general-purpose_2020}). In this process, only the downstream models need updating, thus eliminating the need for extensive re-tuning of the upstream model for different tasks.


In this paper, we focus on G-BLUM systems for its efficiency, and make \textit{three} contributions. 
\textit{First}, while G-BLUM has shown success in platforms like search engines and e-commerce services, their applicability to IM apps remains largely uncharted and requires dedicated investigation due to distinct platform differences. For one, IM apps involve sustained and dynamic user interactions spanning minutes to hours~\cite{grinter2003wan2tlk}, in contrast to the short and focused interactions of search engines~\cite{jansen2022measuring}, or occasional visits to e-commerce platforms~\cite{lee2017relationships}. 
For another, user intentions vary widely on IM apps, from casual chats to urgent communications, unlike the clearer intentions on e-commerce platforms and search engines~\cite{jansen2008determining}. We thus bridge this literature gap through rigorous experiments and evaluations with Snapchat data. 

\textit{Second}, we implement a customized \textit{Transformer}-based~\cite{vaswani_attention_2017} user model for Snapchat users, with two training objectives: \textit{Masked Behavior Prediction} and \textit{User Contrastive Learning} (\S~\ref{sec:training}). 
Furthermore, as users remain engaged, their behavior sequences naturally extend and can therefore exceed the maximum length encountered during training. Making inferences for such sequences remains a challenge in G-BLUM systems.
To overcome this, we utilize a novel position encoding method: Attention with Linear Biases (ALiBi), which performs efficient extrapolation of varying-length sequences~\cite{press_train_2022}. This marks an improvement over previous studies, where fixed input sequence lengths were used (e.g., ~\cite{tao_log2intent_2019, chen_predictive_2018, zhang_general-purpose_2020}),

\textit{Lastly}, we introduce three downstream tasks unseen in previous G-BLUM studies: Reported Account Prediction, Ad View Time Prediction, and Account Self-deletion Prediction. These concern user safety, engagement and churn, three pivotal topics in user research.

\section{Related Work}
\label{sec:relatedwork}
Yang et al. (2017)~\cite{yang_personalizing_2017} applied the word2vec algorithm~\cite{mikolov_distributed_2013} to behavior sequences of Adobe Photoshop users. This research was extended by Tao et al. (2019)~\cite{tao_log2intent_2019} with an encoder-decoder model. 
Chen et al. (2018)~\cite{chen_predictive_2018} used a \textit{recurrent neural network} (RNN) to model user behavior sequences from commercial websites.
More recent G-BLUM studies adopt the \textit{Transformer} architecture~\cite{vaswani_attention_2017,purificato2024user}, which excels due to its self-attention mechanisms, which can capture nuanced interplay between behavior and context and address computational issues like vanishing gradients~\cite{pascanu_difficulty_2013}, making them better suited to encode long, complex sequences.
For instance, Zhang et al. (2020)~\cite{zhang_general-purpose_2020} modeled mobile phone users from their mobile app usage sequences (e.g., app installation, uninstallation) with two training objectives: Sequence Reconstruction and Masked Behavior Prediction and three evaluation tasks: Next Week's App Installation Prediction, Look-alike Audience Extension and Feed Recommendation. Chu et al. (2022)~\cite{chu_simcurl_2022} modeled users of professional design software from their command sequences with contrastive learning and evaluated on predicting user responses to customer surveys. 


In our paper, we also adopt the Transformer architecture. However, we base the design and evaluation of our model on our own proposed conceptualization of G-BLUM systems (\S~\ref{sec:conceptualization_gblum}), resulting in different training objectives and evaluation choices. We do not aim to compare ours with previous G-BLUM systems. 

\section{Conceptualizing G-BLUM Systems}
\label{sec:conceptualization_gblum}
Previous work~\cite[e.g.,][]{zhang_general-purpose_2020, chu_simcurl_2022} considers G-BLUM systems as BLUM systems that learn user representations generalizable to distinct downstream tasks. To guide our model design and evaluation, we break down this definition into five criteria for G-BLUM systems:

\begin{enumerate}
    \item The user model is solely trained on user behavioral data.
    \item The user model's training objective is not tied directly to specific downstream tasks.
    \item The user representations encode behavioral information.
    \item The user representations capture user-specific information that can distinguish one user from another.
    \item The user model can perform distinct downstream tasks.
\end{enumerate}

Criterion 1 concerns that fact that user logs typically contain noisy events unrelated to user actions (e.g., app notifications, error reports), which should be left out.
Criterion 2, 3 and 4 concern the choice of model training objectives. \textit{Masked Behavior Prediction} and \textit{User Contrastive Learning} are two suitable ones.
The former involves randomly masking parts of users' behavior sequences, compelling the model to predict the masked behaviors based on their context and thus learns user behavioral information (fulfilling Criterion 3).
The latter uses a contrastive loss function~\cite{gao-etal-2021-simcse} to maximize the distance between representations of different users and minimize the distance between representations of the same user based on behavioral sequences from different time points. Hence, the model learns user-specific information that distinguishes one user from another (fulfilling Criterion 4). Since these two objectives are not tied to any downstream goal, Criterion 2 is also fulfilled.
Criterion 5 concerns model evaluation, emphasizing that the learned user representations should generalize to different downstream tasks. To this end, we introduce three novel, distinct downstream tasks: Reported Account Prediction, Ad View Time Prediction, and Account Self-deletion Prediction. They concern predicting accounts/users who get reported by other users (e.g., for displaying malicious behaviors), who engage with an ad above a certain duration threshold, and who voluntarily delete their own accounts.

\section{Experiments}

\subsection{Snapchat and Data}
\label{sec:data}
Snapchat is a multimedia IM platform with millions of daily active users and shares similar features (e.g., chats, camera, stories, videos) with other IM platforms~\cite{snapchat_2023}, thus making it a suitable case for exploring the applicability of G-BLUM systems to IM apps. We identify 577 unique user behaviors from raw event logs (e.g., sending a chat, viewing a video, swiping up, and using a filter). 

\paragraph{Training Dataset} Our training data is a random sample of 766,170 U.S. users \textit{active} between April 1, 2023 and April 14, 2023. "Active" is defined as using Snapchat for cumulatively more than 1 minute during this period. The median length of user sequences is 466. 

\paragraph{Evaluation Datasets} We construct datasets for 6 evaluation tasks. See \S~\ref{sec:evaluation} for task and dataset descriptions. Note that users and dates do not overlap between training and evaluation datasets. 

\subsection{Transformer-based User Modeling}
\label{sec:training}
Our user model consists of 12 Transformer Encoder layers with a hidden size of 768 and 12 attention heads per layer. It is trained with a batch size of 512 and a learning rate of $4e\texttt{-}4$ for 20 epochs. 95\% user sequences are used for training while the remaining 5\% is for validation. Training a model takes about 8 hours on 8 Nvidia V100 GPUs.
For training, we truncate each behavioral sequence to maximum 128 tokens, as our preliminary analyses show that longer sequences increase computational overhead while providing minimal additional benefit for evaluation performance. 

\paragraph{Training Objectives} 
During training, the model extracts a random pair of non-overlapping sequences from each user, performs Masked Behavior Prediction on each sequence independently (with 15\% random masking and cross-entropy loss $\mathcal{L}_\textsc{MBP}$), and performs User Contrastive Learning within each mini-batch of behavior sequences. Specifically, 
let $\mathcal{B} = \{ (s_i, s_{i^+}) \}_{i=0}^n$ define a mini-batch of $n$ behavior sequences, where $(s_i, s_{i^+})$ are the sequence pairs extracted from the same user's behavioral logs. The contrastive loss on each sequence pair $(s_i, s_{i^+})$ is 
$
\ell^{i,i^+} =  
-\log \frac{\exp (\mathrm{sim}(e_i, e_{i^+}) / \tau)}
{\sum_{j \neq i} \exp(\alpha_{ij} \cdot \mathrm{sim}(e_i, e_j) / \tau)} \; 
$, where $e_i$ and $e_{i^+}$ are the vectors of sequences $s_i$ and $s_{i^+}$ that belong to user $i$, $e_j$ is the vector of sequence $s_j$ from user $j$, and $\tau$ is the temperature parameter. We obtain each user vector $e$ by aggregating all the behavior/token vectors in the last layer of the user model. The User Contrastive Learning loss over the entire batch is calculated as $\mathcal{L}_\textsc{UCL} = \frac{1}{n} \sum_i \ell^{i,i^+}$. The total combined loss is $\mathcal{L} = \mathcal{L}_\textsc{MBP} + \mathcal{L}_\textsc{UCL}$.

\paragraph{ALiBi}
ALiBi adds a linear bias (a function of the distance between query and key tokens) to the attention scores in a Transformer model~\cite{press_train_2022}. Specifically, let $\mathbf{q}_i$ be the $i$-$th$ query in the input sequence of length $L$, and $\mathbf{K}$ be the key matrix. The attention score of query $i$ is then calculated as: $\texttt{softmax}(\mathbf{q}_i\mathbf{K} + m * [-(i-1), ..., -2, -1, 0, -1, -2, ..., -(L-1-i)])$, where $m$ is a fixed head-specific constant. Following \cite{press_train_2022, zhou2023dnabert2}, we set the values of different $m$ as a geometric sequence (\textit{i.e.,} $\frac{1}{2^1}, \frac{1}{2^2}, ..., \frac{1}{2^n}$). Thus, attention between distant tokens gets penalized, preventing the model from attending to tokens outside of its context window and this helping to generalize to sequences longer than seen previously.

\paragraph{Ablation Analysis} We also perform an ablation study on the impact of the training objectives and ALiBi on our user model.

\subsection{Baselines and User Representation Extraction Methods}
\label{sec:baselines}
We compare our model with several baseline models, including:
\begin{itemize}
  \item Term Frequency (\textbf{TF}) and TF-Inverse Document Frequency (\textbf{TF-IDF}).
  \item Skip-Gram with Negative Sampling (\textbf{SGNS})~\cite{mikolov_distributed_2013}, similar to the word2vec approach in Yang et al. (2017)~\cite{yang_personalizing_2017}. 
  \item \textbf{Untrained} user representations, where a fixed vector is randomly generated for each unique user behavior. This "high-dimensional" TF approach has shown competitive performances in various downstream applications~\cite{arora-etal-2020-contextual}.
  \item Transformer Encoder (\textbf{Enc}) and Decoder (\textbf{Dec}) are our implementation of \texttt{BERT-base} \cite{devlin2018bert} and \texttt{GPT2-117M} \cite{gpt2} for user modeling. We train them respectively with Masked Behavior Prediction and Next Behavior Prediction objective. 
\end{itemize}

\paragraph{User Representation Extraction} TF and TF-IDF are fixed-length vectors that summarize user behavior on the sequence level and are thus directly used as user representations. In contrast, our user model and the other baseline models learn vector representations on the token/behavior level. Therefore, pooling the behavior-level vectors to a sequence-level vector is necessary. We experiment with four pooling strategies, including mean pooling, max pooling, weighted mean pooling, and weighted max pooling. We then select the best performing pooling strategy for each of the user models.

\begin{figure*}[h]
\begin{subfigure}{.25\textwidth}
  \centering
		\begin{tikzpicture}
		\draw (0,0 ) node[inner sep=0] {\includegraphics[width=\linewidth, trim={0cm 0cm 0cm 0cm}, clip]{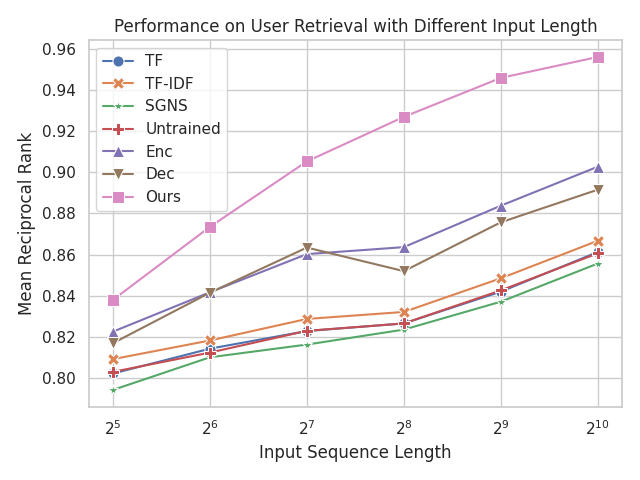}};
		\end{tikzpicture}
		\caption{}
		\label{fig:user_retrieval}
\end{subfigure}%
\begin{subfigure}{.75\textwidth}
		\centering
		\begin{tikzpicture}
		\draw (0,0 ) node[inner sep=0] {\includegraphics[width=\linewidth, trim={0cm 0cm 0cm 0cm}, clip]{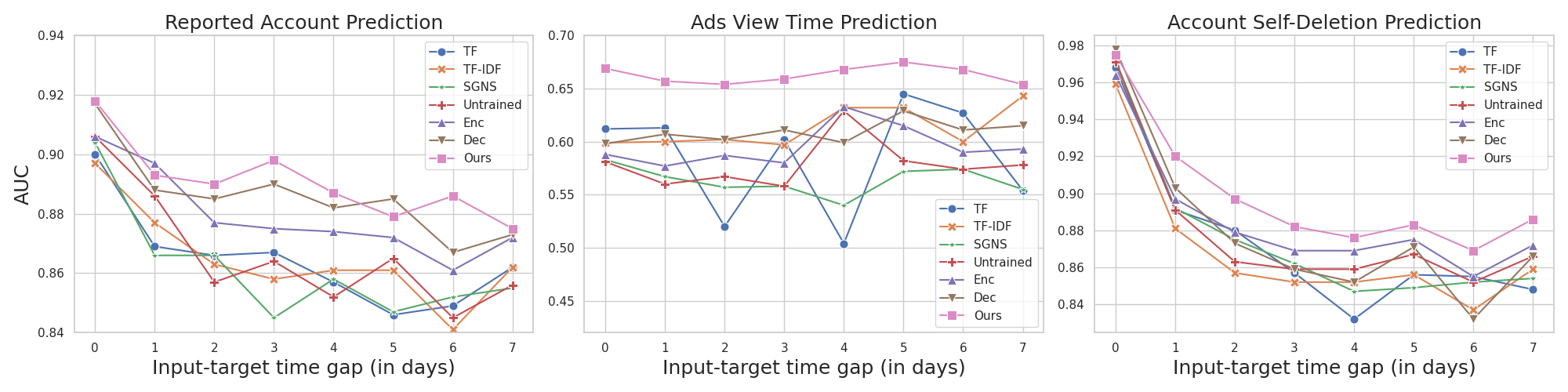}};
		\end{tikzpicture}
		\caption{}
		\label{fig:downstream}
\end{subfigure}
\caption{Model Performance on (a) User Retrieval across Input Sequences Lengths and (b) Three Downstream Tasks across time gaps.}
\label{fig:performance}
\end{figure*}

\subsection{Model Evaluation}
\label{sec:evaluation}
Following Criterion 3 and 4, we expect user representations from our model to encode both behavioral and user-specific information. For the former, we check whether the model can accurately predict a masked behavior given its context (i.e., \textit{Masked Behavior Prediction}). For the latter, we check whether the model can 1) generate similar representations for the same user based on the user's behavioral sequences from non-overlapping periods and dissimilar representations for different users (i.e., \textit{User Representation Similarity Analysis}), and 2) among behavioral sequences of different users, retrieve the two that belong to the same user (i.e., \textit{User Retrieval}). 

\paragraph{Masked Behavior Prediction} We use a random sample of 23,596 users, randomly mask behaviors in the user sequences and check how accurately the model can predict those masked behaviors. To account for varying frequencies of behaviors, we also used a stratified masking procedure, masking each behavior equally frequently. 

\paragraph{User Representation Similarity
Analysis.}  
We use a random sample of 2,001 users and check whether randomly chosen sequences from the same user would be more similar (based on cosine similarity) than those from different users. Larger differences are preferred. 

\paragraph{User Retrieval}
Using the same previous dataset, we assess, given a random user's sequence and a sample of sequences from both that user and other users, whether the same user's sequence can be retrieved. 
Specifically, each sample consists of 101 user sequences: 1 \textit{query} and 100 \textit{candidates}. Within the 100 candidates, there is 1 \textit{positive candidate} that belongs to the query user and 99 \textit{negative candidates} that belong to other users. For each sample, we obtain a vector representation for each behavior sequence, and rank the 100 candidates based on their cosine similarity with the query. We measure the models' performance with Mean Reciprocal Rank (MRR). 
Additionally, we vary the input sequence length at inference time. For baseline models with an input length limitation, we split each behavior sequence into non-overlapping 128-length segments, obtain a vector representation for each segment, and take the average of these representations as the final user representation.

\paragraph{Reported Account Prediction, Ad View Time Prediction, Account Self-deletion Prediction}
We formulate these downstream tasks as binary classification problems. We extract user representations using the latest 128 behaviors, and fit a multi-layer perception (MLP) classifier using random search with 5-fold cross validation for hyperparameter selection. We use AUC as the evaluation metric. In addition, we vary the gap between the user behavior sequences and the labels from 0 to 7 days. 
For these three tasks, we collected balanced datasets of sample sizes between 10,000 and 13,000.

\subsection{Results}
\label{sec:results}

\subsubsection{Masked Behavior Prediction}
Table~\ref{tab:mask} summarizes the performance of our user model (\textit{Ours}) and the baseline model (\textit{Enc}) on Masked Behavior Prediction across the two masking strategies (i.e., \textit{Random} and \textit{Stratified}). We exclude the other baselines, as they do not have a Masked Behavior Prediction training objective. Our model achieves substantially higher prediction accuracy than \textit{Majority Vote} for both masking strategies. Remarkably, even with stratified masking, our model's performance goes down by only about 5 percentage points. While Enc has the best performance for both masking strategies, this is to be expected, as Enc is optimized for this task. That our model performs on par with Enc demonstrates its ability to encode user behaviors. 

\begin{table}[h]
  \centering
  \begin{minipage}[t]{0.45\linewidth}
    \small
    \caption{Accuracy on Masked Behavior Prediction.}
    \label{tab:mask}
    \resizebox{3.5cm}{!} {
    \begin{tabular}{lcc}
      \toprule & Random & Stratified \\
      \midrule
      \textbf{Majority Vote} & 0.2450 & 0.063 \\
      \textbf{Enc} & 0.9379 & 0.8908 \\
      \textbf{Ours} & 0.9221 & 0.8693 \\
    \bottomrule
    \end{tabular}
    }
  \end{minipage}
  \hfill
  \begin{minipage}[t]{0.45\linewidth}
  \small
    \caption{Within- vs Between-user Cosine Similarity.}
    \label{tab:cosine}
    \centering
    \resizebox{4cm}{!} {
    \begin{tabular}{lccc}
      \toprule
              \textbf{Model} & \textbf{Within}   & \textbf{Between}   & \textbf{Difference} $\uparrow$ \\
      \midrule
      \textbf{TF} &  0.7607 & 0.5610 & 0.1997  \\
      \textbf{TF-IDF} & 0.6572 & 0.3811 & 0.2762  \\
      \textbf{Untrained} &  0.7563 & 0.5540 & 0.2023 \\
      \textbf{SGNS} &  0.8880 & 0.7908 & 0.0972 \\
      \textbf{Enc} &0.8046 & 0.6350 & 0.1696 \\
      \textbf{Dec} & 0.6761 & 0.4589 & 0.2172 \\
      \midrule
      \textbf{Ours} & 0.5960 & 0.1910 & \textbf{0.4050} \\
    \bottomrule
  \end{tabular}
  }
  \end{minipage}
\end{table}

\subsubsection{User Representation Similarity Analysis}
Table~\ref{tab:cosine} shows the results from User Representation Similarity Analysis. The \textit{Within} column is the cosine similarity between representations of the same user, while \textit{Between} column is the cosine similarity between representations of different users. 
Because both within-user and between-user cosine similarity scores can be high due to overlapping user behaviors, we focus on the difference between \textit{within-user} and \textit{between-user} cosine similarity (i.e., the \textit{Difference} column), which better reflects the ability of user representations to encode user-specific information and make one user's representation distinguishable from another's. Remarkably, our user model outperforms the best baseline approach (TF-IDF) by about 13 percentage points. 
BERT (Enc) and GPT2 (Dec), however, have led to cosine differences even lower than simple approaches like TF-IDF and Untrained; thus, naively applying BERT/GPT2 to user modeling should be avoided.

\subsubsection{User Retrieval}


Figure \ref{fig:user_retrieval} summarizes results on the User Retrieval task, varying the length of input user sequences from 32 to 1024 by a progression ratio of 2. 
Our model consistently and substantially outperforms all baselines, with performance steadily improving as the input sequence length increases. This shows that our model benefits from longer user sequences and, thanks to ALiBi, can effectively handle longer sequences than seen during training.

\subsubsection{Downstream Task Performance}
Figure \ref{fig:downstream} shows the results on the three downstream tasks across different \textit{time gaps}. 
We highlight two observations. \textit{First}, our model consistently outperforms all baselines (except for time gap 1 and 5 in Reported Account Prediction). 
\textit{Second}, our model can detect malicious accounts (reported by other users) and predict user account self-deletion with high AUC scores up to one week in advance. However, our model predicts ad view time less well (still better than the baselines and chance). This is expected, as ad view time also depends on other important factors like ad length, content and users' cognitive states.

\subsubsection{Ablation Analysis}
Table~\ref{tb:ablation} shows the impact of User Contrastive Learning (UCL) and ALiBi on our model's performance. Without UCL, our model improves slightly on Masked Behavior Prediction, but significantly underperforms other tasks. Also, without ALiBi, performance suffered across all evaluation tasks. 

\begin{table}[h]
\small
	\caption{
		Ablation Study on Masked Behavior Prediction (MBP) objective, User Contrastive Learning (UCL) objective, and ALiBi, evaluated on 6 tasks: MBP, User Representation Similarity Analysis (URSA), User Retrieval (UR), Reported Account Prediction (RAP), Ad View Time Prediction (AVTP), and Account Self-deletion Prediction (ASP). ACC (accuracy), COS (cosine difference), MRR (Mean Reciprocal Rank), and AUC are converted to the same scale.
	}\label{tb:ablation}
	\centering
        \resizebox{6cm}{!} {
	\begin{tabular}{lrrrrrrr}
		\toprule
		 & \textbf{MBP} & \textbf{URSA} & \textbf{UR} & \textbf{RAP} & \textbf{AVTP} & \textbf{ASP}  \\
   		 & ACC & COS & MRR & AUC & AUC & AUC &  \\
		\midrule
		\textbf{Our Model} & 92.21 & 40.50 & 90.77 & 89.07 & 66.30 & 89.85  \\
            ~ - MBP & N/A & -0.27 & -1.28 & +0.23 & -0.80 & -1.53 \\
		~ - UCL &  +1.61  & -28.38 & -4.66 & -0.34 & -3.98 & -0.49  \\ 
		~ - ALiBi & -0.47 & -1.05 & -0.43 & -1.01 & -3.52 & -1.53  \\
            ~ - UCL \& AliBi & +1.58 & -23.54 & -4.53 & -1.15 & -6.76 & -1.34 
 \\
		
		\bottomrule
	\end{tabular}
 }
\end{table}

\section{Conclusion}
\label{sec:conclusion}
In this paper, we demonstrate the use of G-BLUM systems for IM apps using Snapchat data.
We highlight three findings. \textit{First}, our user model achieves high prediction performance on two important downstream domains: user safety and user churn. 
\textit{Second}, sequences as short as consisting of only 128 events can already result in effective general-purpose user representations for IM platforms. G-BLUM thus has the potential to replace traditional user modeling approaches that rely on months worth of (sensitive) user data. 
\textit{Third}, ALiBi can effectively overcome the challenge of making inferences for sequences longer than seen at training time. 

\bibliographystyle{ACM-Reference-Format}
\bibliography{sigir2024}
\end{document}